\documentclass{ifacconf}

\usepackage{natbib}        

\usepackage{amsmath} 
\usepackage{amssymb}  
\usepackage{bbm}
\usepackage{color}
\usepackage{graphics}
\usepackage{graphicx}
\usepackage{epsfig}
\usepackage{epstopdf}
\usepackage{amsfonts}
\usepackage{mathtools}
\usepackage{bm}
\usepackage{fancyhdr}
\usepackage{framed}
\usepackage{float}
\usepackage{cite}
\usepackage{comment}
\usepackage{tikz}
\usetikzlibrary{calc,positioning,shapes,shadows,arrows,fit}

\newtheorem{problem}{Problem}

\newtheorem{remark}{Remark}
\newtheorem{definition}{Definition}
\newtheorem{example}{Example}
\newtheorem{assumption}{Assumption}

\usepackage{algorithm}
\usepackage{algpseudocode}
\usepackage{pifont}

\begin{document}
	\begin{frontmatter}
		
		\title{Probabilistic Plan Synthesis for Coupled Multi-Agent Systems\thanksref{footnoteinfo}} 
		
		\thanks[footnoteinfo]{This work was supported by the H2020 ERC Starting Grant BUCOPHSYS, the Swedish Research Council (VR), the Swedish Foundation for Strategic Research (SSF) and the Knut och Alice Wallenberg Foundation.}
		
		\author[First]{Alexandros Nikou} 
		\author[second]{Jana Tumova} 
		\author[First]{Dimos V. Dimarogonas}
		
		\address[First]{ACCESS Linnaeus Center, School of Electrical Engineering and KTH Center for Autonomous Systems, KTH Royal Institute of Technology, SE-100 44, Stockholm, Sweden. \\ E-mail: \{anikou, dimos\}@kth.se}
		\address[second]{School of Computer Science and Communication, KTH Royal Institute of Technology, SE-100 44, Stockholm, Sweden. \\ E-mail: \{tumova\}@kth.se}
		
		\begin{abstract}  
			This paper presents a fully automated procedure for controller synthesis for multi-agent systems under the presence of uncertainties. We model the motion of each of the $N$ agents in the environment as a Markov Decision Process (MDP) and we assign to each agent one individual high-level formula given in Probabilistic Computational Tree Logic (PCTL). Each agent may need to collaborate with other agents in order to achieve a task. The collaboration is imposed by sharing actions between the agents. We aim to design local control policies such that each agent satisfies its individual PCTL formula. The proposed algorithm builds on clustering the agents, MDP products construction and controller policies design. We show that our approach has better computational complexity than the centralized case, which traditionally suffers from very high computational demands.
		\end{abstract}
		
		
	\end{frontmatter}

\section{Introduction}

Cooperative control of multi-agent systems has traditionally focused on designing distributed control laws in order to achieve global tasks such as consensus, formation control, network connectivity and collision avoidance. Over the last few years,  the field of control of multi-agent systems under high-level task specifications has been gaining attention. In this work, we aim to impose specific probability bounds to each robot in order to satisfy one specification task formula. Such formulas can be ``The probability of a robot to periodically survey regions A, B, C, avoid region D is always more than 0.9", or ``The probability of a robot to visit regions A, B, C, in this order is more that 0.8".

Temporal properties for multi-agent plan synthesis under Linear Temporal Logic (LTL) formulas has been considered e.g., in \citep{guo_203_reconfiguration1, guo_2015_reconfiguration, belta_optimality, zavlanos_2016_multi-agent_LTL}. Timed temporal properties given in Metric Temporal Logic (MTL) and Metric Interval Temporal Logic (MITL) have been addressed in \citep{quottrup_timed_automata, frazzoli_MTL, alex_2016_acc, alex_acc_2017, alex_sofie_ifac_2017} respectively.

Most of the existing formal synthesis frameworks are based on the discretization of the agent's motion in a partitioned environment to a finite Transition System (TS) (the process is called abstraction) under the following assumptions. First, the measurements of the current state are accurate. Second, the transition system is either purely deterministic (namely, each control action enables a unique transition) or purely nondeterministic (namely, each control action enable multiple transitions). However, in realistic applications of robotic systems, noisy sensors and actuators can cause both of the aforementioned assumptions to fail. Motivated by this, we aim to model the multi-agent system in a probabilistic way such that the above issues are taken into consideration.

Some recent works model the system in a probabilistic way and imposes high-level specifications, given in Linear Temporal Logic (see e.g., \citep{belta_ifac_2011_undertain_environment, murray_2012_mdp_ltl, belta_2011_mdp_ltl_tac, topcu_2015_mdp_ltl, belta_2015_mdp_ltl_ijrr}). In \citep{liu_2016_mdp_mitl}, the authors modeled the system with an MDP and computed policies such that the satisfaction of a formula given in MITL, is maximized. Other works model the system in a probabilistic way with Markov Decision Processes (MDPs) and introduce high-level tasks in PCTL (see e.g., \citep{belta_2011_modp, belta_2012_tro_pctlalgorithms, lin_2015_counterexample_single, lin_2015_counterexample_multi}). However, all these works are restricted to single agent planning and they are not expendable to multi-agent systems in a straightforward way since in the multi-agent case potential couplings may occur among the agents.

By extending these works to multi-agent systems, in order to develop an approach in which the system noise, model errors and external disturbances is explicitly considered, in this work we consider that each agent is modeled as a MDP and the task specifications are given in PCTL formulas. Motivated by the fact that in real applications, the agents (robots) are required to collaborate with each other to perform a task, we assume that there are agents in the system that are dependent to each other. They need to communicate, collaborate through sharing a common action in order to achieve a desired task. 

The main contribution of the paper is to develop a strategy for controlling a general framework of $N$ individual MDPs with respect to individual agents' specifications given in PCTL formulas. The proposed solution can handle the dependencies between the agents by considering agents clustering and MDP product construction and has provably better complexity than the centralized approach. When applied to robotic systems, our approach provides a framework for multi-robot control from temporal logic specifications with probabilistic guarantees. To the best of the authors' knowledge this is the first work that addresses the cooperative task planning for multi-agent systems under probabilistic temporal logic specifications in the presence of dependencies between the agents.

The paper is divided into five parts. In Section \ref{sec: preliminaries}, notation and preliminaries are given. Section \ref{sec: prob_formulation} provides the modeling of the system and the problem statement. Section \ref{sec:problem_solution} provides the technical details of the solution. Finally, conclusions and future work directions are discussed in Section \ref{sec: conclusions}.

\section{Notation and Preliminaries} \label{sec: preliminaries}

In this section, the mathematical notation and preliminary definitions from probabilistic model checking that are necessary for this paper are introduced.

Denote by $\mathbb{N}$ the set of natural numbers. Given a set $S$, denote by $|S|$ its cardinality and by $2^S$ the set of all its subsets. Denote by $\underset{i=1}{\overset{N}{\bigtimes}} S_i = S_1 \times \ldots \times S_N$ the $n$-th fold Cartesian product of the sets $S_1, \dots, S_N$. 

An \textit{atomic proposition} $\chi$ is a statement that is either True $(\top)$ or False $(\bot)$.

\subsection{Markov Decision Processes} \label{sec:mdp_definitions}

MDPs offer a mathematical framework for modeling systems with stochastic dynamics. These models provide an effective way for describing processes in which sequential decision making is required for a system.

\begin{definition}
A \textit{probability distribution} over a countable set $S$ is a function $\sigma: S \to [0, 1]$ satisfying $\sum_{s \in S}^{}\sigma(s) = 1$. Define by $\Sigma(S)$ the set of all probability distributions over the set $S$.
\end{definition}

\begin{definition} \label{def:dtmc}
A Discrete Time Markov Chain (DTMC) $\mathcal{D}$ is a tuple $(S, s_0, P)$ where: $S$ is a finite set of states; $s_{0} \in S$ is the initial state; $P : S \times S \to [0,1]$ is the transition probability matrix where for all $s \in S$ it holds that $\sum_{s' \in S}^{} P(s,s') = 1$.
\end{definition}
\begin{definition} \label{def:mdp}
A Markov Decision Process (MDP) $\mathcal{M}$ is a tuple $(S, s_0, Act, T)$ where: $S$ is a finite set of states; $s_{0} \in S$ is the initial state; $Act$ is a finite set of actions (controls); $T: S \to 2^{Act \times \Sigma(S)}$ is the transition probability function.
\end{definition}
Denote by $\mathcal{A}(s)$ the set of all actions that are available at the state $s \in S$ and let $\delta(s, \alpha, s') \in [0, 1]$ be the probability of transitioning from the state $s$ to the state $s'$ under the action $\alpha \in \mathcal{A}(s)$. For a state $s \in S$ and an action $\alpha \in \mathcal{A}(s)$, define the set $\text{Post}(s, \alpha) = \{s' \in S: \delta(s,\alpha,s')>0\}$.


The transition probability function $T$ can be represented as a matrix with $\sum_{i = 0}^{|S|-1} |\mathcal{A}(s_i)|$ rows and $|S|$ columns.

An execution of an MPD is represented by a \textit{path}. Formally, an \textit{infinite path} $r$ is a sequence of states of the form: $r = s_0 \overset{\alpha_0}{\longrightarrow} s_1 \overset{\alpha_1}{\longrightarrow} \dots \overset{\alpha_{k-1}}{\longrightarrow} s_k \overset{\alpha_k}{\longrightarrow} s_{k+1} \dots,$
such that $s_k \in S_k, \alpha_k \in \mathcal{A}(s_k)$ and $\delta(s_k, \alpha_{k+1}, s_{k+1}) > 0, \forall k \geq 0$. A \textit{finite path} $\rho = s_0 \overset{\alpha_0}{\longrightarrow} s_1 \overset{\alpha_1}{\longrightarrow} \dots \overset{\alpha_{n-1}}{\longrightarrow} s_n$ is a prefix of an infinite path ending in a state. In case of the actions are not taken into consideration, the infinite and finite run can be written as $r = s_1s_2 \dots s_n \dots$ and $\rho = s_1s_2 \dots s_n$ respectively.  Denote by $|\rho| = n$ the length of the finite path and by $r(k), \rho(k)$ the $k$-th element of the paths $r, \rho$ respectively. The set of all finite and infinite paths are defined by $FPath$ and $IPath$ respectively.

A control policy at each state of an MDP and is formally defined as follows:

\begin{definition} (Control Policy) A \textit{control policy} $\mu: FPath \\ \to Act$ of an  MDP model $\mathcal{M}$ is a function mapping a finite path $\rho = s_0 \overset{\alpha_0}{\longrightarrow} s_1 \overset{\alpha_1}{\longrightarrow} \dots \overset{\alpha_{n-1}}{\longrightarrow} s_n$, of $\mathcal{M}$ onto an action in $\mathcal{A}(s_n)$ and specifies for every finite path, the next action to be enabled. If a control policy depends only on the last state of the finite path $\rho$, then it is called a \textit{stationary policy}.
\end{definition}

Denote by $M$ the set of all control policies. Under a control policy $\mu \in M$, an MDP becomes a DTMC $\mathcal{D}_\mu$ (see Def. \ref{def:dtmc}). Let $IPath_\mu \subseteq IPath$ and $FPath_\mu \subseteq FPath$ denote the set of infinite and finite paths that can be produced under the control policy $\mu$. For each policy $\mu \in M$, a probability measure $Prob_\mu$ over the set of all paths (under the control policy $\mu$) $IPath_\mu$ is induced. A probability measure $Prob_\mu^{fin}$ over the set of paths $FPath_\mu$ for a finite path $\rho$, is defined as: $Prob_\mu^{fin} (\rho) = 1, \text{if} \ |\rho| = 0$ and $Prob_\mu^{fin} (\rho) = P(s_0,s_1)P(s_1,s_2) \ldots P(s_{n-1},s_n), \text{otherwise}$, where $P(s_k, s_{k+1}), k \geq 0$ are the corresponding transition probabilities in $\mathcal{D}_\mu$.

Define also the set of all infinite paths with prefix $\rho$ as:
\begin{equation*}
C(\rho) = \{r \in IPath_\mu : \rho \ \text{is a prefix of r} \}.
\end{equation*}
By invoking theorems from probability theory \citep{marta_2004_book}, we have that:
\begin{equation*}
Prob_\mu(C(\rho)) = Prob_\mu^{fin}(\rho), \forall \rho \in FPath_\mu.
\end{equation*}
where $Prob_\mu^{fin}(\rho)$ as is defined previously.

\subsection{Probabilistic Computational Tree Logic (PCTL)} \label{sec:PCTL}

Probabilistic Computational Tree Logic (PCTL) \citep{hansson_1994_pctl} is used to express properties of MDPs. PCTL formulas can be recursively defined as follows:
\begin{align}
\varphi &:= \top \ | \ \chi \ | \ \neg \varphi \ | \ \varphi_1 \wedge \varphi_2 \ | \ \mathcal{P}_{\bowtie p} [\psi], \hspace{5mm} (state \ formulas) \notag \\
\psi &:= \bigcirc \varphi \ | \ \varphi_1 \ \mathcal{U}^{\leq k} \ \varphi_2, \hspace{25mm} (path \ formulas) \notag 
\end{align}
where $\chi \in Act$ is an action, $\bowtie = \{<, >, \leq, \geq \}$, $p \in [0, 1]$ and $k \in \mathbb{N} \cup \{\infty\}$. In the syntax above, we distinguish between state formulas $\varphi$ and path formulas $\psi$, which are evaluated over states and paths, respectively. A property of a model will always be expressed as a state formula; path formulas only occur as the parameter of the probabilistic path operator $\mathcal{P}_{\bowtie p} [\psi]$. Intuitively, a state $s$ satisfies $\mathcal{P}_{\bowtie p} [\psi]$ (we write $s \models \mathcal{P}_{\bowtie p} [\psi]$) if there exists a control policy $\mu$ under which the probability of all paths starting from $s$ is in the range of the interval $\bowtie p$. 

For path formulas, we allow the ``next" ($\bigcirc$) operator which is true if the state formula $\varphi$ is satisfied in the next state and the ``bounded until" ($\mathcal{U}^{\leq k}$) which is true if $\varphi_2$ is satisfied within $k$ steps and $\varphi_1$ holds up until that point. The unbounded ``until" operator $\mathcal{U}$ is the same as $\mathcal{U}^{\leq k}$ as $k \to \infty$.

\begin{definition} \citep{hansson_1994_pctl} (Semantics of PCTL) For any state $s \in S$, the satisfaction relation $\models$ is defined inductively as follows:
	\begin{align*}
	s &\models \top \Leftrightarrow \forall s \in S, \\
	s &\models \chi \Leftrightarrow \chi \in \mathcal{A}(s), \label{eq:modified_sem} \\
	s &\models \neg \varphi \Leftrightarrow s \not \models \varphi, \\
	s&\models \varphi_1 \wedge \varphi_2 \Leftrightarrow s \models \varphi_1 \ \text{and} \ s \models \varphi_2, \\
	s &\models \mathcal{P}_{\bowtie p} [\psi] \Leftrightarrow Prob_\mu(s, \psi) \bowtie p,
	\end{align*}
where $Prob_\mu(s, \psi)$ denotes the probability that a path starting from the state $s$ satisfies the path formula $\psi$ under the specific control policy $\mu$. Moreover, for any path $r \in IPath$ we have that:
\begin{align*}
r &\models \bigcirc \varphi \Leftrightarrow r(1) \models \varphi, \\
r &\models \varphi_1 \ \mathcal{U}^{\leq k} \ \varphi_2  \Leftrightarrow \exists \ i \leq k, \notag \\
&\hspace{25mm} r(i) \models \varphi_2 \wedge r(j) \models \varphi_1 \ \forall j < i.
\end{align*}
For the operators $\square$ (always) and $\Diamond$ (eventually) it holds that:
\begin{align}
\mathcal{P}_{\bowtie p} \left[ \Diamond^{\leq k} \varphi \right] &=  \mathcal{P}_{\bowtie p} \left[ \top \ \mathcal{U}^{\leq k} \varphi \right], \notag \\
\mathcal{P}_{\bowtie p} \left[ \square^{\leq k} \varphi \right] &=  \mathcal{P}_{\overline{\bowtie} p} \left[ \Diamond^{\leq k} \neg \varphi \right], \notag
\end{align}
where $\bowtie = \{<, >, \leq, \geq \}$ and $\overline{\bowtie} = \{>, <, \geq, \leq\}$.
\end{definition}

\begin{remark}
Traditionally, the PCTL semantics are defined over a set of atomic propositions. However, in this paper, we aim to introduce dependencies over the actions among the agents. Therefore, the PCTL semantics are defined over a set of actions.
\end{remark}

\subsection{Probabilistic Verification} \label{sec:probabilistic_model_checking}

There are three problems that have generally been considered in probabilistic model checking of stochastic systems:
\begin{itemize}
	\item (Model Checking) Given an MDP $\mathcal{M}$ and a property $\varphi$, check which of the states of the MDP $\mathcal{M}$ satisfy $\varphi$.
	
	\item (Controller Synthesis): Given an MDP $\mathcal{M}$ and a property $\varphi$, find all the control policies under which the formula is satisfied.
	\item (Existence) Given an MDP $\mathcal{M}$ and a property $\varphi$, find, if it exists, a control policy $\mu$ such that the MDP $\mathcal{M}$ satisfies the property $\varphi$ under $\mu$.
\end{itemize}

In this paper, we are mainly interested in the Controller Synthesis problem. The motivation for that is the following: if one control policy fails to guarantee the satisfaction of a formula, it may exists another policy under which the formula is satisfied. We refer the reader to \citep{marta_2004_book, alvaro_1995_probabilistic_model_checking} for more details about probabilistic model checking.

\section{Problem Formulation} \label{sec: prob_formulation}

\subsection{System Model and Abstraction}

Consider a multi-agent team with $N \geq 2$ agents operating in the bounded workspace $\mathcal{W} \subseteq \mathbb{R}^n$. Let $\mathcal{V} = \left\{ 1,\ldots, N\right\}$ denote the index set of the agents. The workspace $\mathcal{W} = \mathop{\bigcup} \limits_{\ell \in \mathcal{W}}^{} \gamma_{\ell}$ is partitioned using a finite number (assume $W$) of regions of interest $\gamma_1, \ldots, \gamma_W$. Denote by $\gamma_\ell^i$ the fact that the agent $i$ is occupying the region $\gamma_\ell$, where $i \in \mathcal{V}, \ell \in \mathcal{W}$.  

We assume that each agent is programmed with a small set of feedback control primitives, which are not completely reliable, allowing it to move inside each region and from a region to an adjacent regions. It is also assumed that the probabilities of these transitions are known.

\begin{assumption} \label{assumption:basic_assumption}
We assume here that an abstraction of the dynamics of each robot into a MDP is given, and that a low level continuous time controller that allows each robot to transit from one region $\gamma_\ell$ to an adjacent region $\gamma_l$ with $\ell,l \in \mathcal{W}$, can be designed.
\end{assumption}

This modeling has been also considered in \citep{belta_2012_tro_pctlalgorithms}.

\begin{definition} \label{def:agent_mdp}
The motion of each agent $i \in \mathcal{V}$ in the workspace can be described by a Markov Decision Process (MDP) $\mathcal{M}_i =(S_i, s_0^i, Act_i, T_i)$ where:
\begin{itemize}
	\item $S_i = \left\{ \gamma_1^i, \gamma_2^i, \ldots, \gamma_W^i \right\}$ is the set of states of agent $i$. The number of states for each agent is $|S_i| = W$, meaning that $S_i$ includes all regions within $\mathcal{W}$.
	\item $s_0^i \in S_i$ is the initial state of agent $i$ (the initial region where agent $i$ may start). Note that the initial state is known and deterministic i.e. we know exactly the region from which each agent starts its motion.
	\item $Act_i$ is a finite set of actions (controls). 
	\item $T_i: S_i \to 2^{Act_i \times \Sigma(S_i)}$ is the transition probability function.
\end{itemize}
\end{definition}
\begin{remark} \label{remark:agents_synch}
We investigate here, under which conditions two or more agents are visiting simultaneously a specific region of the workspace. Consider the $\{i_1,\dots,i_c\} \subseteq \mathcal{V}, c \geq 2$ agents of the multi-agent system. Let $r_{i_1} = s_0^{i_1} s_1^{i_1} s_2^{i_1} \dots s_k^{i_1} \dots s^{i_1}_n, \dots, r_{i_c} = s_0^{i_c} s_1^{i_c} s_2^{i_c} \dots s_k^{i_c} \dots s^{i_c}_n$, be the finite paths of length $n$ that are executed by the corresponding MDPs $\mathcal{M}_1, \dots, \mathcal{M}_c$, respectively, where $s_z^{j} \in S_z, \forall z \geq 0, j \in \{i_1,\dots,i_c\}$. Then, if there exists $k \geq 1$ such that for all the $k$-th elements of the above runs (in the same position at every path) it holds that $s_k^{i_1}=\dots=s_k^{i_c} = s^{\text{meet}}_k$, then we say that the agents $\{i_1,\dots,i_c\}$ are visiting simultaneously the same region $s_k$. If there does not exist such region $s^{\text{meet}}_k$, then the agents cannot meet simultaneously to one region.
\end{remark} 

\subsection{Handshaking Actions}

The motivation for introducing dependencies in the multi-agent system comes from real applications where more than one agents (robots) need to collaborate with each other in oder to perform a desired task. For example, imagine two aerial manipulators that are required to meet and grasp an object simultaneously and deliver it to a specific location in a warehouse.

In order to be able to introduce dependencies in the actions between the agents, we write the action set of each agent as: $Act_i = \{\Pi_i, \hat{\Pi}_i\}, i \in \mathcal{V}$, where $\Pi_i$ is a finite set of actions that the agent $i$ need to execute in collaboration with other agents (\textit{handshaking} actions) and $\hat{\Pi}_i$ is a finite set of actions that the agent $i$ executes independently of the other agents (\textit{independent} actions). For the independent actions it holds that: $\hat \Pi_i \cap \hat \Pi_j = \emptyset, \forall i \neq j, i,j \in \mathcal{V}$.

The independent actions can always be executed without any constraints. On the other hand, for the handshaking actions, we have the following requirements:
\begin{itemize}
    \item First, the agents are required to meet and occupy the same region of the workspace (not necessarily a specific region).
	\item Once they meet, they need to execute \text{simultaneously} the same action.
	\item All the agents that share an action are required to execute it in order for the task to be completed properly.
\end{itemize}
Formally, the handshaking actions are defined as follows:
\begin{definition} \label{def:handshaking_actions}
(Handshaking Actions) Let $\{i_1, \dots, i_{c}\} \subseteq \mathcal{V}, c \geq 2$ be a set of agents that need to collaborate in order to execute simultaneously a task under the action $\alpha$. The following two properties should hold in order for $\alpha$ to be well-posed handshaking action:
\begin{enumerate}
\item $\alpha \in \bigcap_{i \in \{i_1,\dots,i_c\}} \Pi_{i}$.
\item Let the following finite paths of length $n$: 
\begin{align*}
r_{i_1} &= s_0^{i_1} \overset{\alpha_0^{i_1}}{\longrightarrow} \ldots  \overset{\alpha_{k-1}^{i_1}}{\longrightarrow} s_{k}^{i_1} \overset{\alpha}{\longrightarrow} s_{k'}^{i_1} \dots \overset{\alpha_{n-1}^{i_1}}{\longrightarrow} s_n^{i_1}, \\
&\qquad \qquad \vdots \\
r_{i_c} &= s_0^{i_c} \overset{\alpha_0^{i_c}}{\longrightarrow} \ldots \overset{\alpha_{k-1}^{i_c}}{\longrightarrow} s_{k}^{i_c} \overset{\alpha}{\longrightarrow} s_{k'}^{i_c} \dots \overset{\alpha_{n-1}^{i_c}}{\longrightarrow} s_n^{i_c},
\end{align*}
be executed by the MDPs $\mathcal{M}_{i_1}, \dots, \\ \mathcal{M}_{i_c}$ respectively. Here, $s_{k}^{i_1}, \dots, s_{k}^{i_c}$ are the regions that the $i_1, \dots, i_c$ should occupy in order to execute the handshaking action $\alpha$ simultaneously. Then, there should exist at least one $k \geq 0$ such that $s_{k}^{i_1} = \dots = s_{k}^{i_c} = s^{\text{meet}}_k$ and $\delta(s_{k}^{j}, \alpha, s_{k'}^{j}) > 0$ for at least one $s_{k'}^j \in \text{Post}(s_k^j,\alpha)$ for every $j \in \{i_1,\dots,i_c\}$.
\end{enumerate} 
\end{definition}
Notice that the same condition for a state $s^{\text{meet}}_k$ as in condition (2) was mentioned in Remark \ref{remark:agents_synch}, but here the existence of a common action $\alpha$ is also required. It should be also noted that every region of the workspace in which the agents can potentially meet, can serve as a region that a handshaking action can be executed (if such an action exists).

\subsection{Dependencies} \label{sec:dependencies}

Suppose that one agent $i$ receives a cooperative task that involves other agent's $j \in \mathcal{V} \backslash \{i\}$ participation. This means that both agents need to execute the same action 
at the same region so as for the task to be performed. The dependencies are formally defined as:
\begin{definition} \label{def:dependency_relation}
The agents $i,j \in \mathcal{V}$ are called \textit{dependent} if one the following statements holds:
\begin{enumerate}
	\item Agent $i$ depends on agent $j$ if $\Pi_{i} \cap \Pi_j \neq \emptyset$.
	\item Agent $i$ depends on agent $j$ if $\Pi_{i} \cap \Pi_j \neq \emptyset$.
	\item There exist at least one region $s^{\text{meet}}_k, k \geq 0$ of the workspace such that the second condition of Def. \ref{def:handshaking_actions} holds.
\end{enumerate}	
\end{definition}
Conditions (1),(2) can be checked by comparing all the elements of the sets $\Pi_{i}, \Pi_j, \forall i,j \in \mathcal{V}$ one by one. Condition (3) can be checked by using graph search algorithms. 

\begin{remark}
It should be noticed from the above definitions that all the agents that share an action, they are required to meet and execute it simultaneously.
\end{remark}

\begin{remark}
Due to fact that the control policies are defined over finite paths, the handshaking actions are defined with respect to finite paths as well. Therefore, the graph search algorithm for condition (3) is searching into a finite graph.
\end{remark}

\begin{assumption} \label{assumption:dependency_assumption}
It is assumed that there exists at least $2$ agents that are dependent. Otherwise, there exist no dependencies between the agents and the problem that is later defined can be straightforwardly solved by solving the controller synthesis methodology of Section \ref{sec:problem_solution} for each agent independently.
\end{assumption}

\subsection{Problem Statement}

We define here the problem that we aim to solve in this paper:

\begin{problem} \label{problem:basic_problem}
	Given $N$ agents performing in the workspace $\mathcal{W}$, under the Assumptions \ref{assumption:basic_assumption}, \ref{assumption:dependency_assumption}, individual task specification formulas $\varphi_1, \ldots, \varphi_N$ over the actions $\Pi_i \cup \hat{\Pi}_i, i \in \mathcal{V}$ given in PCTL with semantics as in Sec. \ref{sec:PCTL}, synthesize individual control policies $\mu_1, \ldots, \mu_N$ (if there exists one) which guarantee the satisfaction of the formulas $\varphi_1, \ldots, \varphi_N$ respectively.
\end{problem}

\section{Problem Solution} \label{sec:problem_solution}

\subsection{Overview}

An overview of the proposed solution is given as follows:

\begin{itemize}
	\item Step 1: First, the dependencies among the agents are modeled as a dependency graph (see Section \ref{section:dependency_graph}). The agents are split into clusters and each cluster contains the agents that are dependent according to Def. \ref{def:dependency_relation}.
	\item Step 2: For each cluster of agents, the mutual specification $\varphi_m$ and the product MDP $\widetilde{M}$ are defined (see Section \ref{sec:product_MDP}). 
	\item Step 3: By utilizing the controller synthesis algorithms of Section \ref{sec:algorithms}, we design a control policy $\widetilde{\mu}$ of each cluster that guarantees the satisfaction of $\varphi_m$ (if such a control policy exists). We provide in Sec. \ref{sec:path_policy_projection} the definition of successful control policies, which project onto local control policies $\mu_1, \dots, \mu_N$ for each agent, which finally are a solution to Problem 1.
\end{itemize}	
An algorithm describing all the steps of the proposed procedure is given in Section \ref{sec:algorithm}. Probabilistic model checking algorithms, which can compute all the control policies under which a PCTL formula $\varphi_{m}$ is satisfied, are presented in Section \ref{sec:algorithms}. The computational complexity of the proposed framework is discussed in Section \ref{section:complexity}.

Problem \ref{problem:basic_problem} can be solved in a centralized way by computing the product of all individual MDPs $\mathcal{M}_i, i \in \mathcal{V}$ (see Definition \ref{def:product_MDP}) and perform the proposed methodology of this paper to the centralized system without any clustering among the agents. However, this solution leads to a high computational burden and state explosion of the product MDP $\mathcal{M}$. A comparison of the computational complexity of the proposed framework that exploits the potential sparsity of dependencies with the centralized approach is discussed in Section \ref{section:complexity}.
	
\subsection{Modeling the Dependencies} \label{section:dependency_graph}

Based on the dependency relation of the Def. \ref{def:dependency_relation}, the dependency graph associated with the handshaking actions $\Pi_i, i \in \mathcal{V}$ is defined as follows:
\begin{definition} \label{def:dependency_graph}
The \textit{dependency graph} $\mathcal{G} = (\mathcal{V}, \mathcal{E})$, is an undirected graph that consists of the set of vertices $\mathcal{V}$ in which each of the agents is node for the graph and the edge set $\mathcal{E}$ which is defined as follows:
\begin{equation*}
\mathcal{E} = \{\{i, j\} : i \ \text{is dependent to} \ j \ \text{and} \ i,j \in \mathcal{V}, i \neq j\}.
\end{equation*}
\end{definition}
In order to proceed, the following definition is required:
\begin{definition} \citep{mesbahi_2010_graph_theory}
Let $\mathcal{G} = (\mathcal{V}, \mathcal{E})$ be an undirected graph. Then every graph $\mathcal{G}' = (\mathcal{V}', \mathcal{E}')$ with $\mathcal{V}' \subseteq \mathcal{V}$ and $\mathcal{E}' \subseteq \mathcal{E}$ if called a \textit{subgraph} of the graph $\mathcal{G}$.
\end{definition}
\begin{definition} \label{def:dependency_cluster}
The set $\mathcal{C} = \{C_\ell : \ell \in \mathbb{M}\} \subseteq \mathcal{V}$, where $\mathbb{M} = \{1,\dots,m\}$, forms a \textit{dependency cluster} if and only if $\forall i,j \in \mathcal{C}$ there is a path from node $i$ to node $j$ in the dependency graph $\mathcal{G}$. 
\end{definition}
Define the function $f: \mathcal{V} \to \mathbb{M}$ which maps each agent to the index of the cluster that it belongs to. It can be observed that $\bigcup_{\ell \in \mathbb{M}} C_\ell = \mathcal{V}$ and $\sum_{\ell \in \mathbb{M}}^{}  |C_\ell| = |\mathcal{V}| = N$. 

Each agent $i \in \mathcal{V}$ for which there exist no $j \in \mathcal{V}$ such that $j \in C_{f(i)}$ will be called an \textit{independent agent}. For an independent agent it holds that $|C_{f(i)}| = 1$.

From Definition \ref{def:dependency_cluster}, it follows that every dependency cluster $C_\ell \in C, \ell \in \mathbb{M}$ is the vertex set of the subgraphs $G^{(\ell)} = (\mathcal{C}_\ell, \mathcal{E}_\ell), \mathcal{E}_\ell \subseteq \mathcal{E}, \ell \in \mathbb{M}$ of the system graph $\mathcal{G}$. 

Loosely speaking, two agents belong to the same cluster when they are directly dependent or transitively dependent by a dependency chain. An example of a dependency graph and dependency clusters is given as follows:
\begin{example} \label{example:dependency_graph_example}
	Consider $N = 6$ agents with $\mathcal{V} = \{1,\ldots,6\}$, $\mathcal{E} = \{\{1, 2\}, \{3, 4\}, \{4, 5\}\}$. The $m = 3$ clusters are given as: $C_1 = \{1,2\}, C_2 = \{3,4,5\}$ and $C_3 = \{6\}$ and the corresponding subgraphs $\mathcal{G}^{(1)} = (C_1, \mathcal{E}_1 = \{1,2\}), \mathcal{G}^{(2)} = (C_2, \mathcal{E}_2 = \{\{3,4\}, \{4,5\}\})$ and $\mathcal{G}^{(3)} = (C_3, \mathcal{E}_3 = \emptyset)$. Moreover, $f(1) = f(2) = 1, f(3) = f(4) = f(5) = 2, f(6) = 3$. The dependency graph is depicted in Fig. \ref{fig:dependency_graph_example}.
	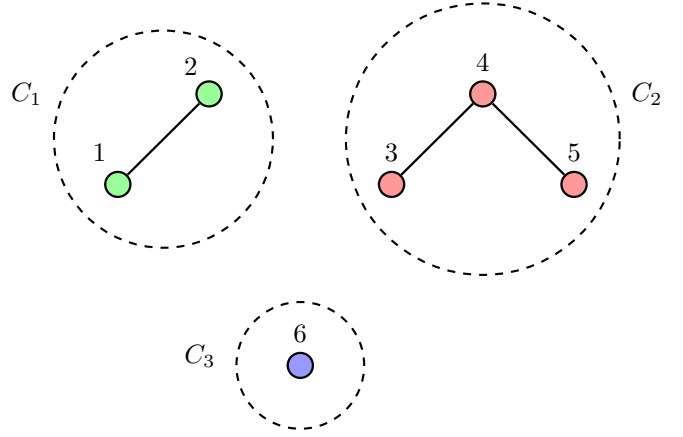
\begin{figure}[ht!]
		\vspace{2mm}
		\centering
		\begin{tikzpicture}    [scale = 1.2]    
		\node(1) [line width = 0.85] at (-3,0)[shape=circle,draw][fill=green!40] {$ $};
		\node(2) [line width = 0.85] at (-2,1)[shape=circle,draw][fill=green!40] {$ $};
		\node(3) [line width = 0.85] at (0,0)[shape=circle,draw][fill=red!40] {$ $};
		\node(4) [line width = 0.85] at (2,0)[shape=circle,draw][fill=red!40] {$ $};
		\node(5) [line width = 0.85] at (1,1)[shape=circle,draw][fill=red!40] {$ $};
		\node(6) [line width = 0.85] at (-1,-2)[shape=circle,draw][fill=blue!40] {$ $};
		
		\path [-] [line width = 0.85]
		(1) edge node [above]      {}   (2)
		(3) edge node [above]      {}   (5)
		(4) edge node [above]      {}   (5);
		
		\draw[black,thick,dashed] (-2.5,0.5) circle (1.2cm);
		\draw[black,thick,dashed] (1,0.5) circle (1.5cm);
		\draw[black,thick,dashed] (-1,-2) circle (0.7cm);
		
		\node at (-3.2, 0.35) {$1$};
		\node at (-2.2, 1.3) {$2$};
		\node at (0, 0.35) {$3$};
		\node at (1, 1.35) {$4$};
		\node at (2, 0.35) {$5$};
		\node at (-1, -1.65) {$6$};
		
		\node at (-4, 1.0) {$C_1$};
		\node at (2.8, 1.0) {$C_2$};
		\node at (-2.1, -1.9) {$C_3$};
		\end{tikzpicture}
		\caption{An example of a dependency graph $\mathcal{G} = (\mathcal{V}, \mathcal{E})$ and its subgraphs for $N = 6$ agents.}
		\label{fig:dependency_graph_example}
	\end{figure}
\end{example}
According to the mathematical derivation above, Assumption \ref{assumption:dependency_assumption} is modified as follows:
\begin{assumption}
There exists at least one dependency cluster $C_\ell, \ell \in \mathbb{M}$ (as was defined in Definition \ref{def:dependency_cluster}) of the dependency graph $\mathcal{G}$ of the under consideration multi-agent system, which contains at least two dependent agents. I.e., there exists $\ell \in \mathbb{M}$ such that: $|C_\ell| = 2, \text{if} \ N = 2$ and $|C_\ell| \in [2, N-1], \text{if} \ N > 2$.
\end{assumption}

By employing the above computation, the initial multi-agent system is modeled as $m$ subgraphs $\mathcal{G}^{\ell}, \ell \in \mathbb{M}$ which capture the dependencies between the agents, as they are defined in Def. \ref{def:dependency_relation}. This forms a convenient modeling of the system's dependencies in order to compute the product MDP of every subsystem $\ell \in \mathbb{M}$ in the next Section.

\subsection{Product Markov Decision Process} \label{sec:product_MDP}

Define here the \emph{mutual specification} of a cluster of agents $C_\ell$ as:
\begin{align} 
\varphi_m^{\ell} = \bigwedge_{i \in C_\ell}^{} \varphi_i, \ell \in \mathbb{M}, \label{eq:formula_mut}
\end{align}
over the set of actions $\bigcup_{i \in C_\ell}^{} \left(\Pi_i \cup \hat{\Pi}_i\right)$. If the satisfaction of $\varphi_m^\ell$ for each cluster $C_\ell$ is guaranteed, it holds by definition that the satisfaction of all the individual formulas $\varphi_i, i \in \mathcal{V}$ is guaranteed as well. Thus, a method for finding a team control policy that guarantees the satisfaction of $\varphi_m^\ell, \ell \in \mathbb{M}$ should be provided.

In the sequel, we construct a product MDP that captures the collaborative behavior of all the agents within a cluster. Having $\widetilde{M}_\ell$, allow us to synthesize a control policy $\widetilde{\mu}$ for $C_\ell$, which guarantees the satisfaction of the collaborative formula $\varphi_m^\ell$. Subsequently, the team control policy $\widetilde{\mu}_\ell$ can be projected onto the local agents' control policies $\mu_1, \dots, \mu_N$ which are a solution to Problem 1.

\begin{definition}\label{def:product_MDP} (Product MDP) The \textit{product MDP} $\widetilde{\mathcal{M}}_\ell$ for the cluster of agents $C_\ell$ is a tuple $(\widetilde{S}_\ell, \widetilde{s}_{0}^{\ell}, \widetilde{Act}_\ell, \widetilde{T}_\ell)$ where:
\begin{itemize}
	\item $\widetilde{S}_\ell = \underset{i \in C_\ell}{\overset{}{\bigtimes}} S_i$ is the set of states. 
	\item $\widetilde{s}_{0}^{\ell} = \underset{i \in C_\ell}{\overset{}{\bigtimes}} s_0^i$ is the initial state.
	\item $\widetilde{Act}_\ell = \bigcup_{i \in C_\ell}^{} Act_i = \bigcup_{i \in C_\ell}^{} \left\{\Pi_i, \hat{\Pi}_i \right\}$ is the set of actions.
	\item $\widetilde{T}_\ell : \widetilde{S}_\ell \to 2^{\widetilde{Act}_\ell \times \Sigma(\widetilde{S}_\ell)}$ is the transition probability function for the product system. Similar to $\delta$ of Def. \ref{def:mdp}, we define $\widetilde{\delta}(\widetilde{s}, \widetilde{\alpha}, \widetilde{s}') \in [0,1]$ the probability of transitioning from the state $\widetilde{s}$ to the state $\widetilde{s}'$ under the action $\widetilde{\alpha}$. Let $C_\ell = \{i_1, \dots, i_{|C_\ell|}\}$ be an enumeration of the agents of the cluster $C_\ell$. Then, $\widetilde{\delta}_\ell$ is defined as follows:
	\begin{enumerate}
		\item $\widetilde{\delta}((s_{i_1}, \ldots, s_{i_{|C_\ell|}}), \alpha, (s'_{i_1}, \ldots, s'_{i_{|C_\ell|}})) = \\ \prod_{j \in C_\ell}^{} \delta_{j}(s_{j}, \alpha, s'_{j})$, if $\alpha \in \bigcap_{j \in C_\ell}^{} \mathcal{A}(s_{j})$.
		\item $\widetilde{\delta}( (\widetilde{s}_{i_1}, \ldots, \widetilde{s}_{k_1}, \ldots, \widetilde{s}_{k_\nu}, \ldots, \widetilde{s}_{i_{|C_\ell|}}), \alpha, (\widetilde{s}_{i_1}, \\ \ldots, \widetilde{s}'_{k_1}, \ldots, \widetilde{s}'_{k_\nu}, \ldots, \widetilde{s}_{i_{|C_\ell|}})) = \prod_{j = 1}^{\nu} \delta_{k_j}(s_{k_j}, \alpha, \\ s'_{k_j})$ if $$\alpha \in \left[\bigcap_{j=1}^{\nu} \mathcal{A}(s_{k_j}) \right] \mathbin{\Bigg\backslash} \left[\bigcup_{z \in C_\ell \backslash \{k_1, \ldots, k_\nu\}}^{} \mathcal{A}(s_z) \right],$$ for $k_j \in C_\ell, j \in \{1, \ldots, \nu\}$.
	\end{enumerate}
\end{itemize}
\end{definition}

Intuitively, (1) denotes that all the agents $i_1, \dots, i_{|C_\ell|}$ of the cluster $|C_\ell|$ are located in the states $s_{i_1}, \ldots, s_{i_{|C_\ell|}}$ respectively, and they are simultaneously transiting to the states $s'_{i_1}, \ldots, s'_{i_{|C_\ell|}}$ with action $\alpha$. (2) denotes that among all the agents of the cluster $C_\ell$, only the agent $\{k_1, \dots, k_\nu\} \subsetneq C_\ell$ are transiting simultaneously to the states $s'_{k_1}, \dots, s'_{k_\nu}$ respectively. (2) can not be handshaking action since for the handshaking action all the agents of the cluster should transit simultaneously to the next state. In order for (1) to be a handshaking transition according to Def. \ref{def:handshaking_actions} it is required also that $s_{i_1} = \ldots = s_{i_{|C_\ell|}}$.

\begin{remark}
In the case of a cluster $\ell \in \mathbb{M}$ that contains an independent agent $i \in \mathcal{V}$ with the property $|C_\ell| = |C_{f(i)}| = 1$, the product MDP $\widetilde{\mathcal{M}}_\ell$ coincides with the individual MDP $\mathcal{M}_i$ of Def. \ref{def:agent_mdp} ($\widetilde{\mathcal{M}}_\ell \equiv \mathcal{M}_i$).
\end{remark}
The infinite path $\widetilde{r}$, the finite path $\widetilde{\rho}$, the control policy $\widetilde{\mu}$  and the set of all infinite and finite paths $\widetilde{FPath}$ and $\widetilde{IPath}$, are defined similarly to Sec. \ref{sec:mdp_definitions}. 

\subsection{Designing the Control Policies $\mathcal{\widetilde{\mu}}$} \label{sec:path_policy_projection}

The product MDP $\widetilde{M}_\ell, \ell \in \mathbb{M}$ of each cluster captures the paths and the control policies of the agents that belong to the same cluster and they are required to collaborate for achieving a task or acting independently. 

By employing the controller synthesis algorithms (see Section \ref{sec:algorithms}), the control policies $\widetilde{\mu}_\ell$ for the team of agents in each cluster can be designed. The algorithms can compute all the control policies $\widetilde{\mu}_\ell$ that guarantees the satisfaction of formula $\varphi^\ell_m$ from \eqref{eq:formula_mut}. 

What remains is to project these policies onto the individual control policies of the agents of each cluster in such a way that they serve as a solution to Problem 1.

Consider a cluster of agents $C_\ell = \{i_1, \dots, i_{|C_\ell|}\}$. A control policy $\widetilde{\mu}_\ell(\widetilde \rho) = \widetilde{\mu}_\ell(\widetilde{s}_0\widetilde{s}_1 \dots \widetilde{s}_n) \subseteq \widetilde{Act}$ for the finite path $\widetilde{\rho} = \widetilde{s}_0\widetilde{s}_1 \dots \widetilde{s}_n$ of length $n$, where $\widetilde{s}_k = (s^k_{i_1}, \dots, s^k_{i_{|C_\ell|}}), k \in \{1,\dots,n\}$, projects onto the local individual control policies $\mu_{j}(s^1_{j}, \dots, s^n_{j}), j \in \{i_1, \dots, i_{|C_\ell|}\}$, of the agents $\{i_1, \dots, i_{|C_\ell|}\}$ of the cluster $C_\ell, \ell \in \mathbb{M}$. Note that: $\mu_{j} \subseteq \widetilde{Act} \ \bigg\lvert_{j} \subseteq Act_{j} , j \in \{i_1, \dots, i_{|C_\ell|}\}$, and $\widetilde{Act} \ \bigg\lvert_{j}$ is the set of actions of the agent $j$ that are appearing in the s et $\widetilde{Act}$. 

The set $\widetilde{\mu}(\widetilde{\rho})$ contains control policies that are either handshaking or independent. Let us also define the following set of handshaking actions: $\text{Succ}(\alpha, \ell) = \{\alpha \in \Pi_{i_1} \cap \dots \cap \Pi_{i_{|C_\ell|}}: \alpha \in \widetilde{\mu}_\ell(\widetilde{\rho})\}$, which is the subset of $\widetilde{\mu}(\widetilde{\rho})$ that contains the handshaking actions. We need to search now if all the projections $\mu_{i_j}, \forall j \in \{1,\dots,|C_\ell|\}$ follow the handshaking rules of Def. \ref{def:handshaking_actions}.

\begin{definition} \label{def:control_policy_solution}(Successful Control Policy) 
Let $\widetilde{\mu}_\ell(\widetilde \rho) = \widetilde{\mu}_\ell(\widetilde{s}_0\widetilde{s}_1 \dots \widetilde{s}_n) \subseteq \widetilde{Act}$ be a control policy of a cluster $C_\ell$. The control policy $\widetilde{\mu}_\ell(\widetilde{\rho})$ is called \textit{successful} if for all $\alpha \in \text{Succ}(\alpha,\ell)$ it holds that $s^n_{i_1} = \dots = s^n_{i_{|C_\ell|}}$ and $\delta(s_j^n, \alpha, (s_j^n)') > 0$ 
for at least one $(s_j^n)' \in \text{Post}(s_j^n,\alpha), j \in \{i_i, \dots, i_{|C_\ell|}\}$.
\end{definition}

Let $SP(\ell) = \left\{\widetilde{\mu}_\ell(\widetilde{\rho}) \subseteq \widetilde{Act}: \widetilde{M}_\ell \models \varphi_m^\ell \right\}, \ell \in \mathbb{M},$ denotes the set of all the control policies that guarantee the satisfaction of the formula $\varphi_m^\ell$. All the control policies of $SP$ needs to be checked if they are successful. If $SP(\ell) = \emptyset$ for at least one $\ell \in \mathbb{M}$, then the Problem \ref{problem:basic_problem} has no solution. The set $SP(\ell)$ is computed by employing the algorithms of Section \ref{sec:algorithms}.

\subsection{Proposed Algorithm} \label{sec:algorithm}

The proposed procedure of solving Problem \ref{problem:basic_problem} can be shown in Algorithm \ref{alg:basic_algorithm}. The function $\text{checkDepend}()$ determines the dependent agents according to Def. \ref{def:dependency_relation}. The product and projection that were introduced in Sec. \ref{sec:product_MDP}, \ref{sec:path_policy_projection}, are computed by the functions $\text{product}(), \text{projection}()$ respectively. The algorithms of Sec. \ref{sec:algorithms} are incorporated in the function $\text{controlSynthesis}()$. By employing Def. \ref{def:control_policy_solution}, the function $\text{succPolicy}()$ determines if a sequence of control policies are successful.

\begin{remark}
	Even though our proposed solution is centralized in each cluster, it is partially decentralized in terms of the whole multi-agent system.
\end{remark}

\begin{algorithm}
	\caption{- SolveProblem1($\cdot$)}
	\begin{algorithmic}[1]
		\State \textbf{Input:} MDPs: $\mathcal{M}_1,\dots,\mathcal{M}_N$; \\
		\hspace{11mm} PCTL Formulas: $\varphi_1,\dots,\varphi_N$
		\State \textbf{Output:} $\mu_1, \dots, \mu_N$ \\
		\State $\mathcal{C} = \{C_\ell, \ell \in \mathbb{M}\}=\text{checkDepend}(Act_1, \dots, Act_N)$
		\State $\varphi_m^{\ell} = \bigwedge_{i \in C_\ell}^{} \varphi_i$
		\For {$z \in C_\ell =\{i_1, \dots, i_{|C_\ell}|\}$}
		   \State $\widetilde{M}_\ell = \text{product}(\{\mathcal{M}_j, j \in C_\ell\})$
		   \State $SP(\ell) = \text{controlSynthesis}(\widetilde{M}, \varphi_m^\ell)$
		\For {$\widetilde{\mu}_\ell \in SP(\ell)$}
		   \State $\{\mu_1,\dots,\mu_N\} = \text{projection}(\widetilde{M}_\ell)$
		   \If {$\text{succPolicy}(\{\mu_1,\dots,\mu_N\}) = \top$}
		      \State $\text{solFound} = 1$
		      \State  $\text{return} \ \{\mu_1,\dots,\mu_N\}$ \Comment{Solution found}
		   \Else
		       \State $\text{go to} \ 12$ \Comment{Search other control policies}
		   \EndIf
		\EndFor 
		   \If {$\text{solFound} \neq 1$}
		       \State Problem 1 has no solution
		   \EndIf
		\EndFor
	\end{algorithmic}
	\label{alg:basic_algorithm}
\end{algorithm}

\subsection{Algorithms for Probabilistic Control Synthesis} \label{sec:algorithms}

We are investigating here algorithms of computing all the control policies $\widetilde{\mu}_\ell \in SP(\ell)$. Once these control policies are found, then by following Algorithm \ref{alg:basic_algorithm}, the individual policies $\mu_{j}, j \in \{i_1, \dots, i_{|C_\ell|}\}$ can be designed and the Problem 1 is solved (if there exists a solution). For more details about the algorithms we refer to \citep{marta_2004_book, marta_2007_stochastic, marta_paper_2_2011, alvaro_1995_probabilistic_model_checking, belta_2012_tro_pctlalgorithms}

First, define $\text{Sat}(\varphi^\ell_{m}) = \{s \in S: s \models \varphi^\ell_m\}$ as the set of states that satisfy $\varphi_m^{\ell}$. Then we have: $\text{Sat}(\top) = S, \text{Sat}(\pi) = \{s \in S : \pi \in \mathcal{A}(s)\}, \text{Sat}(\neg \varphi^\ell_{m}) = S \backslash \text{Sat}(\varphi^\ell_{m}), \text{Sat}(\varphi^{\ell}_{m,1} \wedge \varphi^{\ell}_{m,2}) = \text{Sat}(\varphi^{\ell}_{m,1}) \cap \text{Sat}(\varphi^{\ell}_{m,2})$ for two PCTL formulas $\varphi^{\ell}_{m,1}, \varphi^{\ell}_{m,2}$. Define also the minimum and the maximum probabilities of satisfying the formula under the control policy $\mu$ for a starting state $s$:
\begin{subequations}
	\begin{align}
	Prob_{\max}(s, \psi) &= \sup_{\mu \in M} \{Prob_\mu(s,\psi)\}, \label{eq:prob_max} \\
	Prob_{\min}(s, \psi) &= \inf_{\mu \in M} \{Prob_\mu(s,\psi)\}. \label{eq:prob_min}
	\end{align}
\end{subequations}
where $M$ is set of all control policies. It has been proved in \citep{alvaro_1995_probabilistic_model_checking},  that the model checking problem problem of the operator $\mathcal{P}_{\bowtie p}[\psi]$ can be reduced to the computation of \eqref{eq:prob_max}, \eqref{eq:prob_min} according to the following:
\begin{align}
\bullet \ \text{If} \ \bowtie &= \{\geq, >\} \ \text{then} \notag \\ 
&\hspace{15mm} s \models \mathcal{P}_{\bowtie p}[\psi] \Leftrightarrow Prob_{\min}(s, \psi) \bowtie p. \label{eq:condition_p_min} \\
\bullet \ \text{If} \ \bowtie &= \{\leq, <\} \ \text{then} \notag \\ 
&\hspace{15mm} s \models \mathcal{P}_{\bowtie p}[\psi] \Leftrightarrow Prob_{\max}(s, \psi) \bowtie p. \label{eq:condition_p_max} 
\end{align}

For the controller synthesis (as was defined in Section \ref{sec:probabilistic_model_checking}) of the path operators $\mathcal{P}_{\bowtie p}[\bigcirc \varphi^{\ell}_{m}]$, $\mathcal{P}_{\bowtie p}[\varphi^{\ell}_{m} \mathcal{U}^{\leq k} \varphi^{\ell}_{m}]$, we utilize the Algorithms 2,3 respectively as follows:

\subsubsection{Algorithm 2}

If the formula is $\varphi^{\ell}_{m} = \mathcal{P}_{\bowtie p}[\bigcirc \varphi^{\ell}_{m,1}]$, initially the maximum probability of satisfying $\varphi^{\ell}_{m}$ at the state $s\in S$:
\begin{equation}
Prob_{\max}(s, \varphi^{\ell}_{m}) = \underset{\alpha \in \mathcal{A}(s)}{\text{max}} \left(\sum_{s' \in Sat(\varphi^{\ell}_{m,1})}^{} \delta(s, \alpha, s')\right), \label{eq:alg_1}
\end{equation}
is computed for every $s \in S$. By replacing $\max$ with $\min$ in \eqref{eq:alg_1}, $Prob_{\min}(s, \bigcirc \varphi^{\ell}_{m,1})$ can be computed. Define the vector $\Phi(s) = 1, \text{if} \ s \in Sat(\varphi^{\ell}_{m, 1})$ or $\Phi(s) = 0, \text{otherwise}$. Perform now the matrix multiplication $X = T \cdot \Phi$. $X$ is a vector whose entries are the probabilities of satisfying $\bigcirc \varphi_1$, where each row corresponds to a state-action pair. After obtaining the vector $X$, eliminate the state-actions pairs whose probabilities are not in the range of $\bowtie p$ by taking into consideration the conditions \eqref{eq:condition_p_min}, \eqref{eq:condition_p_max}. This operation determines all the states $s \in S$ and all the actions $\mu \in M$ that satisfy the formula $\varphi^{\ell}_{m}$.

\subsubsection{Algorithm 3}

For the formula of the form $\phi_m^{\ell} = \mathcal{P}_{\bowtie p} \left[\varphi^{\ell}_{m,1} \ \mathcal{U}^{\leq k} \ \varphi^{\ell}_{m,2}\right]$, define by $S^{\text{yes}} = \text{Sat}(\varphi^{\ell}_{m,2}), S^{no} = S \backslash \left[\text{Sat}(\varphi^{\ell}_{m,1}) \cup \text{Sat}(\varphi^{\ell}_{m,2})\right],$ and $S^{rem} = S \backslash (S^{yes} \cup S^{\text{no}})$ the states that always satisfy the specification, the states that  never satisfy the specification and the remaining states, respectively. Compute the maximum probability of satisfying $\varphi^{\ell}_{m}$ at the state $s\in S$ as: $Prob_{\max}(s, \varphi^{\ell}_{m}) = 1 \ \text{or} \ 0, \text{if} \ s \in S^{\text{yes}} \ \text{or} \ s \in S^{\text{no}}$ respectively. For $s \in s \in S^{\text{rem}}$ and $k \ge 0$ compute recursively the following: 
\begin{align}
&Prob_{\max}(s, \varphi^{\ell}_{m}, k) \notag \\ 
&= \underset{\alpha \in \mathcal{A}(s)}{\text{max}} \left( \sum_{s' \in S^{rem}}^{} \delta(s, \alpha, s') Prob_{\max}(s, \varphi^{\ell}_{m}, k-1)\right. \notag \\
&\hspace{40mm}\left. +\sum_{s' \in S^{yes}}^{} \delta(s, \alpha, s')\right), \label{eq:prob_k}
\end{align}
with $Prob_{\max}(s, \varphi^{\ell}_{m}, 0) = 0$. The computation can be carried out in $k$ iterations, each similar to the process of Algorithm 2. By replacing $\max$ with $\min$ in \eqref{eq:prob_k}, $Prob_{\min}(s, \varphi^{\ell}_{m}, k)$ can be computed.

\subsubsection{Algorithm 4}

The form $\phi_m^{\ell} = \mathcal{P}_{\bowtie p} \left[\varphi^{\ell}_{m,1} \ \mathcal{U}^{} \ \varphi^{\ell}_{m,2}\right]$ is in fact the same as $\phi_m^{\ell} = \mathcal{P}_{\bowtie p} \left[\varphi^{\ell}_{m,1} \ \mathcal{U}^{\leq k} \ \varphi^{\ell}_{m,2}\right]$ as $k \to \infty$. With this approach, Algorithm 3 can be used to solve this problem.

\begin{remark}
The resulting control strategies of the aforementioned algorithms are stationary. Therefore, the control policies $\widetilde{\mu}_\ell(\widetilde{s}_1\widetilde{s}_2\dots\widetilde{s}_n)$ depend only to the state $\widetilde{s}_n$.
\end{remark}
\subsection{Computational Complexity} \label{section:complexity}

According to \citep{katoen}, the model checking of an MDP $\mathcal{M}$ is polynomial in the number of states of $\mathcal{M}$ and linear in the length of the formula $\varphi$. Denote by $|\varphi|$ the length of the formula $\varphi$ in terms of the number of the operator it has e.g., $|\mathcal{P}_{\geq 0.5} [\bigcirc \{red\} ]| = 2$ . The complexity can be expressed mathematically as $$\mathcal{O}(\text{poly}(|\mathcal{M}|) |\varphi| \kappa(\varphi)),$$ where $\kappa(\varphi) = \max \{k: \phi_1 \mathcal{U}^{\leq k} \varphi_2\}$, $\varphi_1, \varphi_2$ are subformulas of $\varphi$ and $\varphi_1 \mathcal{U}^{\leq k} \varphi_2$ are possible until operators involving in $\varphi$. Define also $$\text{poly}(n) = 2^{\mathcal{O}(\log(n))}.$$ If $\varphi$ does not contain a bounded until operator then $\kappa(\varphi) = 1$.

The number of states of the the product MDP in the centralized solution is $|\widetilde{S}| = \prod_{i \in \mathcal{V}} |S_i| = W^N$ and the corresponding complexity is in the class of $$O = \mathcal{O}\left( 2^{\mathcal{O}(N \log(W))} \cdot |\varphi^{\ell}_m| \cdot \kappa(\varphi_{m}^\ell) \right),$$ where $\varphi^\ell_{m}$ as it is defined in \eqref{eq:formula_mut} for $|C_\ell| = N$.

The worst case complexity of the proposed framework is when $1$ agent is independent and the other $N-1$ agents are dependent to each other. Then, there are two clusters $\ell \in \{1,2\}$: the first contains the independent agent and the other one contains the remaining agents. The corresponding MDPs have $|\widetilde{S}_\ell| = |\underset{i \in C_\ell}{\overset{}{\bigtimes}} S_i| = W^{|C_\ell|}, \ell \in \{1,2\}$ states i.e., $W, W^{N-1}$ states respectively. Thus, the worst case complexity of our framework is: 
\begin{align}
&\widetilde{O} = \mathcal{O}\bigg(2^{\mathcal{O}(\log(W))}\cdot |\varphi_{m}^1| \cdot \kappa(\varphi_{m}^1) \notag \\ 
&\hspace{20mm}+ 2^{\mathcal{O}((N-1) \log(W))} \cdot |\varphi_{m}^2| \cdot \kappa(\varphi_{m}^2)\bigg).
\end{align}
The best case complexity of the proposed framework is when every agent is dependent to at most one other agent. Formally, if $N$ is odd number then $|C_{\ell'}| = 1$ for only one $\ell' \in \mathbb{M}$ and $|C_\ell| = 2, \forall \ell \in \mathbb{M} \backslash \{\ell'\}$. In this case, the best case complexity is in the class: 
\begin{align}
&\bar{O} = \mathcal{O}\bigg( \sum_{\ell \in \mathbb{M} \backslash \{\ell'\}}^{} [2^{\mathcal{O}((N-1) \log(W))} \cdot |\varphi_{m}^\ell| \cdot \kappa(\varphi_{m}^\ell)] \notag \\ 
&\hspace{30mm}+ 2^{\mathcal{O}(\log(W))}  \cdot |\varphi_{m}^{\ell'}| \cdot \kappa(\varphi_{m}^{\ell'}) \bigg). \notag
\end{align}
If $N$ is even number, then previous summation in performed in all the elements $\ell \in \mathbb{M}$. In total, it holds that: $\bar{O} < \widetilde{O} < O$ which verifies that our proposed framework achieves significantly better computational complexity than the centralized one. 

\section{Conclusions and Future Work} \label{sec: conclusions}

We have proposed a systematic method for designing control policies for multi-agent systems. We assume that the system is under the presence of model uncertainties and actuation failures, thus the modeling is performed through MDPs. The agents are divided into dependency clusters which indicate the team of agents that they need to share an action in order to achieve a desired task. With the proposed framework, each agent is guaranteed to perform a task given in PCTL formulas. The computational complexity of the proposed framework is significantly better than the complexity of the centralized framework. Future efforts will be devoted towards performing the abstraction of the stochastic system which is given according to Assumption \ref{assumption:basic_assumption}.

\bibliographystyle{plainnat}
\bibliography{references}   
\end{document}